\documentclass[epj]{webofc}
\usepackage[varg]{txfonts}   

\newcommand{\ket}{\rangle}
\usepackage{multirow}
\wocname{EPJ Web of Conferences}
\woctitle{INPC 2013}
\begin{document}
\title{Electromagnetic Reactions and Few-Nucleon Dynamics}
%
%

\author{Sonia Bacca\inst{1}\fnsep\thanks{\email{bacca@triumf.ca}}
}

\institute{TRIUMF, 4004 Wesbrook Mall,
Vancouver, B.C. V6J 2A3, Canada
          }

\abstract{%

We present an update on  recent theoretical studies of electromagnetic reactions obtained by using the Lorentz integral transform method. The $^4$He nucleus will be the main focus of this report: results for the photo-disintegration and the electro-disintegration processes will be shown, as well as a recent calculation of polarizability effects in muonic atoms.
 We also discuss the exciting possibility to investigate inelastic reactions for medium-mass  nuclei in coupled-cluster theory, highlighted by the recent application to the $^{16}$O photo-nuclear cross section.
}
\maketitle
\section{Introduction}
\label{intro}
 The study of electromagnetic reactions is fundamental to our understanding of the nuclear dynamics, because a clear comparison between theory and experiment
is facilitated by the perturbative nature of the electromagnetic probe.
By concentrating first  on few-nucleons,  where ab-initio approaches are applicable, one can study the sensitivity of electromagnetic reactions to different Hamiltonians, which are major ingredients in the theoretical calculations. 
For the nucleon-nucleon ($NN$) force several potentials  are available that  fit  
 $NN$ scattering data with high accuracy~\cite{AV18,EM}.
However, since the nuclear potential has an  effective nature, it is in principle a many-body operator; thus,  three-body forces ($3NF$s)~\cite{EHM,UIX}, and in principle higher-body forces, are expected to be play a role.
To understand and the determine the realistic  $3NF$s, one ideally seeks for observables involving at least three nucleons that show sensitivity to the $3NF$. Bound-state observables, such as binding energies~\cite{aaron} and radii~\cite{Maxime} are the simplest examples. Elastic hadronic reactions have also been proven sensitive to the $3NF$s~\cite{Nollett, Hupin}.
Electromagnetic inelastic reactions are complementary quantities that do show sensitivity to the nuclear Hamiltonians and are key for a comprehensive understanding of all the facets of the nuclear dynamics.

In this paper, we will present an update on recent studies of electromagnetic reactions analyzed with the Lorentz integral transform method~\cite{LIT}. In Sec.~\ref{sec:1} we  present the main calculational techniques used in our studies.
Applications to electromagnetic reactions with light nuclei and  with the medium-mass nuclei are presented in Sec.~\ref{sec:2}  and in Sec.~\ref{sec:3}, respectively. Finally, in Sec.~\ref{sec:5} we draw our conclusions.

\section{Theoretical Methods}
\label{sec:1}

\subsection{The Lorentz Integral Transform}
\label{sub-sec-1}
We  are interested in electromagnetic induced reactions  
for which a fundamental ingredient is the  nuclear response
function. In the case of an inclusive process it is defined as 
\begin{equation}
\label{resp}
S_{O}(\omega,q)=\int \!\!\!\!\!\!\!\sum _{f} 
\left|\left\langle \Psi_{f}| O(q)| 
\Psi _{0}\right\rangle\right|
^{2}\delta\left(E_{f}-E_{0}-\omega \right)\,. 
\end{equation}
We denote the energy and momentum transferred by the electromagnetic probe to the nucleus with $\omega$ and $q$, while  
$| \Psi_{0/f} \rangle$ and 
$E_{0/f}$ are the initial and final state wave functions and energies. The operator ${O}$
represents a general electromagnetic excitation operator.
From Eq.~(\ref{resp}) it is clear that also energies $E_f$ where the nucleus is broken into many
fragments are involved. Thus, in principle one needs the knowledge of 
all  possible final states $| \Psi_{f} \rangle$ in the continuum.
This fact constitutes a major obstacle, because it is known that the exact knowledge of
the break-up states is limited in mass number and in energy. 
We 
circumvent this problem by using the Lorentz integral transform (LIT) method \cite{LIT}.
This approach is based on introducing an integral transform the of the response
function as
\begin{equation} \label{lorenzo}
  {L}(\omega_0,\Gamma )=\int_{\omega_{\rm th}}^{\infty} d\omega \frac{S_{O}(\omega, q)}{(\omega -\omega_0)
               ^2+\Gamma^2}\:\mbox{,}
\end{equation}   
where $\omega_{\rm th}$ is the threshold energy, $\omega_0$ is the centroid and  $\Gamma > 0 $ is the width of the Lorentzian kernel. 
By using the closure relation one finds
\begin{equation}
\label{lorenzog} { L}(\omega_0, \Gamma)= \langle \psi_0 |
{{O}(q)}^{\dagger}\frac{1}{\hat
  {H}-z^*}\frac{1}{\hat{H}-z}{O}(q)|\psi_0 \rangle = \langle
\widetilde{\Psi}_{z,q}^{O} | \widetilde{\Psi}_{z,q}^{O} \rangle \:\mbox{,} 
\end{equation}
where we introduced the complex energy
$z=E_0+\omega_0+i\Gamma$. The LIT of the response function in Eq.~(\ref{lorenzog})
can be computed directly by solving the Schr{\"o}dinger-like equation
\begin{equation}
({H}-z)|\widetilde{\Psi}_{z,q}^{O}
\rangle={O}(q)|{\Psi_{0}}\rangle\,,\label{liteq}
\end{equation}
where $|\widetilde{\Psi}_{z,q}^{O} \rangle$
is a state with bound-state-like asymptotics. Because of this property one is allowed to use bound-state techniques to
solve Eq.~(\ref{liteq}). So far, hyper-spherical harmonics expansions~\cite{EIHH}, no core shell model~\cite{stetcu2007} and coupled-cluster theory~\cite{cclit}
have been used to solve the LIT equation.
 The response function
 $S_{O}(\omega,q)$ is typically obtained from a
numerical inversion of the integral transform ~\cite{EfL99,andreasi2005}  and is independent
on the choice of the width $\Gamma$.

\subsection{Hyper-spherical Harmonics}
\label{sub-sec-2}
Most of the applications of the LIT method have been so far obtained solving Eq.(\ref{liteq}) 
with hyper-spherical harmonics (HH) expansions in the mass range $3\le A \le 7$.
The HH approach starts from the Jacobi coordinates
\begin{equation}
\label{jacobc}
\boldsymbol {\eta }_{0~~}=\frac{1}{\sqrt{A}}\sum _{i=1}^{A}\mathbf{r}_{i}\,,
\qquad \boldsymbol {\eta }_{k-1}=\sqrt{\frac{k-1}{k}}\left( \mathbf{r}_{k}-\frac{1}{k-1}\sum _{i=1}^{k-1}\mathbf{r}_{i}\right),\, k=2,...,A\,,
\end{equation}
where $\mathbf{r}_i$ are the particle coordinates.  Using the
$\boldsymbol {\eta }_{i}$ one can then transform to hyper-spherical
coordinates composed of one hyper-radial coordinate
$\rho=\sqrt{\sum_{i=1}^{A-1} \boldsymbol {\eta }_{i}^2}$ and a set of
$(3A-4)$ angles that we denote with $\Omega$ (for more details
see~\cite{Nir}). Using this coordinates one can recursively construct
the hyper-spherical harmonics $\mathcal{Y}_{[K]}$ and use them as a complete basis to expand
the wave function. Such expansion reads
\begin{equation}
\label{expans}
\Psi( \boldsymbol {\eta }_{1}, ..., \boldsymbol {\eta }_{A-1},
  s_1,...,s_A, t_1,...,t_A)= \sum_{n}^{n_{\rm max}}  \sum_{[K]}^{K_{\rm max}}C_{[K] n} \, H_{n}(\rho) \,
{\cal Y}_{[K]}(\Omega,s_1,...,s_A, t_1, ...,t_A ),
\end{equation}
where $s_i$ and $t_i$ are the spin and isospin of the nucleon i,
respectively; $C_{[K] n}$ is the coefficient of the expansion, labeled
by $[K]$, which represents a cumulative quantum number that includes
the grand-angular momentum $K$;  $n$ labels the hyper-radial wave function
$H_{n}(\rho)$.  
To further increase the convergence rate of
the calculations, we typically employ an effective interaction
in the  hyper-spherical harmonics (EIHH), as first introduced
in~\cite{EIHH}. 
 Eq.(\ref{liteq}) is solved by expanding its right-hand-side in terms of the HH basis
and then calculating the LIT with the Lanczos algorithm~\cite{Mario}.

\subsection{Coupled-Cluster Theory}
\label{sub-sec-3}

Coupled-cluster (CC) theory~\cite{kuemmel1978,bartlett2007} is a very
efficient bound-state technique, applied with success on several
medium-mass nuclei~\cite{hagen2010b,hagen2012a,hagen2012b,roth2012}. 
 The ground-state of the system is given by $|\psi_0\ket =
\exp(T)|\phi_0\ket$, where $|\phi_0\ket$ is a Slater determinant and
$T$ generates particle-hole ($ph$) excitations. The theory is exact when excitations 
up to A-$ph$ are considered. However, a very efficient approximation scheme is obtained
when $T$ is taken to be the sum of a $1p$-$1h$ operator $T_1$ and a $2p$-$2h$
operator $T_2$.  This goes under the name of coupled-cluster with
singles-and-doubles  excitations~\cite{bartlett2007}. 
 Given the fact the the LIT method requires the solution of a bound-state equation, it is kind of
natural to try solving it within coupled-cluster theory to access the medium-mass nuclei. We develop this new
method based on coupling the LIT with CC in~\cite{cclit}.
The Schr\"{o}dinger-like equation 
(\ref{liteq})   becomes
\begin{equation} 
\label{cc_psi1}
  (\bar{H}-z)|\widetilde{\psi}_R(z)\rangle =
  \bar{O}(q) | 0_R\rangle,
\end{equation}
where $\bar H = \exp(-T)\hat H\exp(T)$ is the similarity transformed
Hamiltonian, and $\bar O(q)=\exp(-T) O(q)\exp(T)$ is
the similarity-transformed excitation operator.
The state $| 0_R \rangle $ is the right ground-state of the 
the non-hermitian  Hamiltonian $\bar H$, which is in general different from
 the left eigenstate
 $\langle 0_L |$.
 Here $|\widetilde{\psi}_R(z)\ket = {R}(z)
|\phi_0\ket$, and $R$ is the excitation operator. The latter is also
expanded in
particle-hole excitations, consistently with the single-double scheme as
\begin{equation}
 R(z) = R_0 + \sum_{i a} R_{i}^{{a}} \hat{c}^\dagger_{{a}}
\hat{c}_{{i}} + {1\over 4}\sum_{i j a b} R_{i j}^{{a b}}
\hat{c}^\dagger_{a}\hat{c}^\dagger_{b} \hat{c}_{j}\hat{c}_{i} 
\;.  
\end{equation}
An equivalent formulation can be derived for the left Schr{\"o}dinger-like equation 
in terms of  $\langle\widetilde{\psi}_L(z)|$ and the left excitation operator $L$. 
The LIT becomes then  $  {L}(\omega_0,\Gamma )=\langle\widetilde{\psi}_L(z)| \widetilde{\psi}_R(z)\ket  $ and can be computed
efficiently by employing a generalization of the Lanczos algorithm for
non-symmetric matrices.

\section{Applications to Few-Nucleon Systems}
\label{sec:2}

The first application of the LIT method we discuss is the total  
 {\bf photo-absorption} cross section $\sigma_{\gamma}$.  The latter is related to the dipole
response function $S_{\rm E1}(\omega)$  by
\begin{equation}
\sigma_\gamma(\omega)=4\pi^{2}\alpha\omega S_{\rm E1}(\omega)\,,
\label{cs}
\end{equation}
\noindent where $\alpha$ is the fine structure constant and $q=\omega$.  The dipole operator is $E1=\sum_i^Z z_i$, where $z_i$
are the coordinates of the Z protons in the center of mass frame. $\sigma_{\gamma}$ has been extensively studied with the LIT method for nuclei with mass number $ 3 \le A \le 7$.
Particular attention has been devoted  to the $^4$He nucleus, both from the theoretical and the experimental
viewpoint.  This stable nucleus is  an ideal testing ground for continuum effects of  the $3NF$s.
 Exchange currents, which can also affect electromagnetic observables, are implicitly
 included in the dipole response function via the Siegert theorem.
\begin{figure}[htb]
\centering
\sidecaption
\includegraphics[width=7cm,clip]{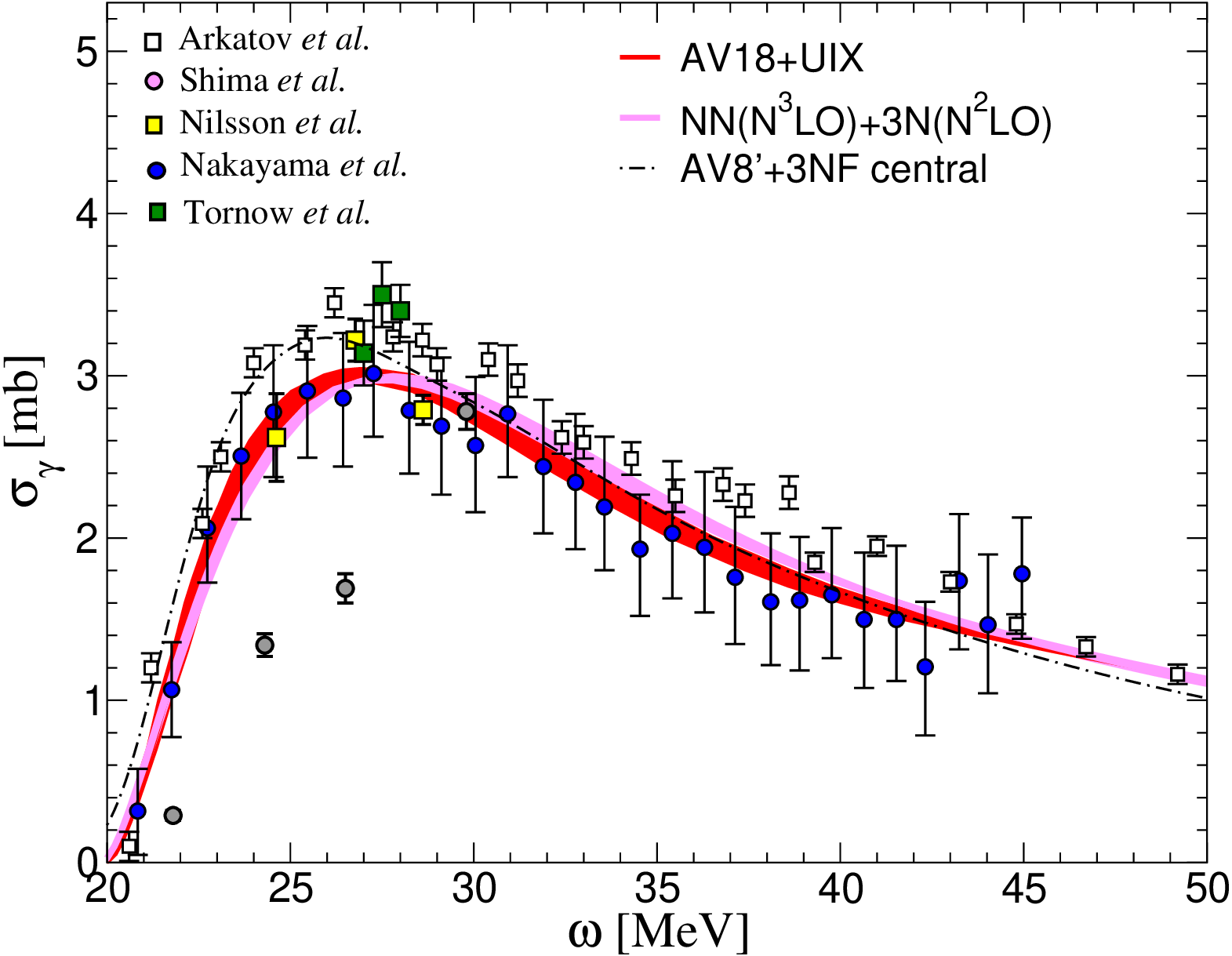}
\caption{ Photo-absorption cross section for $^4$He: 
calculations with the AV18+UIX, with $NN$(N$^3$LO) + $3NF$(N$^2$LO) 
and with a central $3NF$ in comparison with available experimental data
(see Ref.~\cite{PRL_photon} for all experimental reference).}
\label{fig_photon}       
\end{figure}
Calculations of  $\sigma_{\gamma}$ with the AV18+UIX realistic potential were performed in~\cite{PRL_photon}.
Chiral effective field theory forces
(with the $NN$ force at N$^3$LO and the  $3NF$ at N$^2$LO) 
were published
 in  \cite{SofiaPhoton},
where the LIT method was used in conjunction with the 
no core shell model. 
 A very recent calculation was performed using the complex scaling method \cite{Wataru} for a   purely central
three-body force. These theoretical curves are all shown in Fig.~\ref{fig_photon} and form a kind of theoretical (error) band.
 The variation of the cross section with the Hamiltonians  is of the order of $10\%$ at the peak. This is much less than the difference in the available experimental data, also shown in Fig.~\ref{fig_photon}.
In particular, the data from Shima {\it et~al.} \cite{Shima} are a factor of 2 smaller than all other measurements including the most recent ones from TUNL~\cite{TUNL}.
Because the theoretical sensitivity to changes in the Hamiltonian is smaller than the difference in the experiments, it is unfortunately not possible to discriminate among $3NF$s.

The higher accuracy reached by theory with respect to the experiments can be, on the other hand,  of advantage to estimate nuclear polarizability corrections in {\bf muonic atoms} with a better precision.
Stimulated by the spectroscopic measurements of the Lamb shift  2$S$-2$P$ transitions on muonic Hydrogen~\cite{pohl}, which leads to an extraction of the proton radius  deviating 7$\sigma$  from the measurements in ordinary atoms, a series of experiments are planned at PSI to measure the charge radius in muonic Helium. The objective is to investigate whether the discrepancy with respect to ordinary atoms persists or changes for
different charge Z and mass number A.
 The precision of the extracted radius is limited by the accuracy of the
calculated nuclear polarizability corrections.  The leading-order corrections to the Lamb shift energy are related
to integrals of the dipole response function with particular energy weights. In the specific, the leading dipole correction  $\delta^{(0)}_{E1} $
and the leading Coulomb distortion correction $\delta_{C}^{(0)}$ are defined as
\begin{eqnarray}
\label{corr}
\nonumber
\delta^{(0)}_{E1} & =&-\frac{2\pi m_r^3}{9}  (Z\alpha)^5 \frac{9}{4\pi Z^2} \int^\infty_{\omega_{\rm th}}
d\omega \sqrt{\frac{2m_r}{\omega}} S_{E1}(\omega), \\
\delta_{C}^{(0)}& =& -\frac{2\pi m_r^3}{9}  (Z\alpha)^6
             \frac{9}{4\pi Z^2}\int^\infty_{\omega_{\rm th}} d\omega 
\left[\frac{m_r}{\omega} \left(\frac{1}{6}+\ln\frac{2m_r Z^2
    \alpha^2}{\omega}\right) -\frac{17}{16}Z\alpha \left(\frac{2m_r}{\omega}\right)^{3/2}\right]
S_{E1}(\omega),
\end{eqnarray}
where $m_r$ is the reduced mass of the muon-nucleus system.
Thus, one can estimate these corrections  with the $S_{E1}(\omega)$ data extracted from the
photo-absorption cross sections. This strategy was attempted e.g. on $\mu\,^4$He$^+$ by several people~\cite{Bernabeu:1973uf, Rinker:1976en, Friar:1977cf}), leading to a $20\%$ error in the nuclear polarization correction.
Using theory instead, which is more precise than experiment as shown in Fig.~\ref{fig_photon}, we were recently able to
calculate these corrections for  $\mu\,^4$He$^+$ with a much better precision~\cite{chen}. The numbers for  $\delta^{(0)}_{E1} $ and $\delta_{C}^{(0)}$ are shown in Table~\ref{table:polHe4} for two different model spaces in the HH expansion (indicated by $K_{\rm max}$ and $K_{\rm max}-4$),  where  two realistic Hamiltonians are used. As one can see, the numerical error due to the model space truncation is below $1\%$, while the dependence on the potential is roughly $6\%$.  Thus, the overall precision is much better than previous estimates based on experimental data.
\begin{table}[htb]
 \begin{center}
 \caption{Nuclear polarization corrections to the $2S$-$2P$ Lamb
          shift in $\mu\,^4$He$^+$. Numbers are in meV and $K_{\rm max}=22$ and 20 for  AV18+UIX and $NN$(N$^3$LO)+$3N$(N$^2$LO), respectively. } 
 \label{table:polHe4}
 \begin{tabular}{ccc}
 \hline
 correction          & AV18+UIX  & $NN$(N$^3$LO)+$3N$(N$^2$LO) \\
                     & $K_{\rm max}~/~K_{\rm max}-4$ & $K_{\rm max}~/~K_{\rm max}-4$ \\
 \hline
 $\delta^{(0)}_{E1}$  & -4.418 / -4.399$~\;~\;$     & -4.701 / -4.697$\;$    \\ 
$\delta^{(0)}_{C}$    &  0.512 / 0.509$~\;~\;$     &  0.546 / 0.545$\;$    \\
 \hline
 \end{tabular}
 \end{center}
 \end{table}

An other electromagnetic reaction that one can investigate with the LIT method is the inclusive
{\bf electron scattering} off a nucleus. Mass numbers $A=3$ and $A=4$ have been quite extensively studied in
conjunction with the hyper-spherical harmonics expansion. For the $^4$He case we have recently calculated
the (longitudinal) response function $S_L(\omega,{q})$ to the charge 
density operator~\cite{el_PRL,el_PRC}.
 $S_L$ is well known for not being very sensitive
to exchange currents at low $q$, so we focused on a sensitivity study of this observable to the $3NF$s.
We observed that the difference between calculations with $NN$ only and calculations which include $3NF$s is increasing at low $q$. At $q=50$ MeV/c, where no experimental data exist, $3NF$s lead to up to a $50\%$ quenching effect.
A $10\%$ difference in $S_L$ is obtained by using   different three-body Hamiltonians~\cite{sonia_fb}. Thus, future precise experiments can potentially discriminate among realistic potentials.
\begin{figure}
\includegraphics[scale=0.26,clip=]{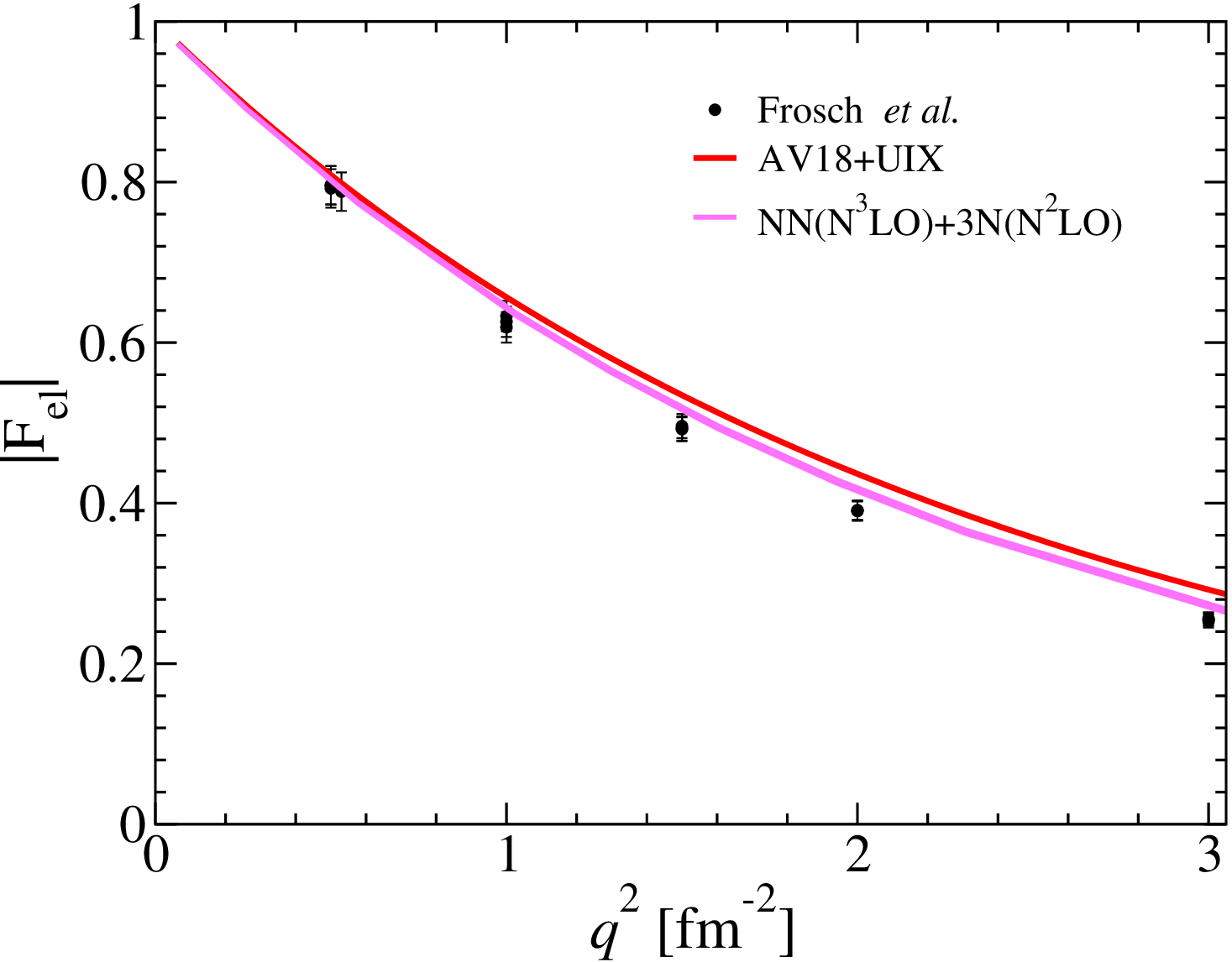}
\includegraphics[scale=0.275,clip=]{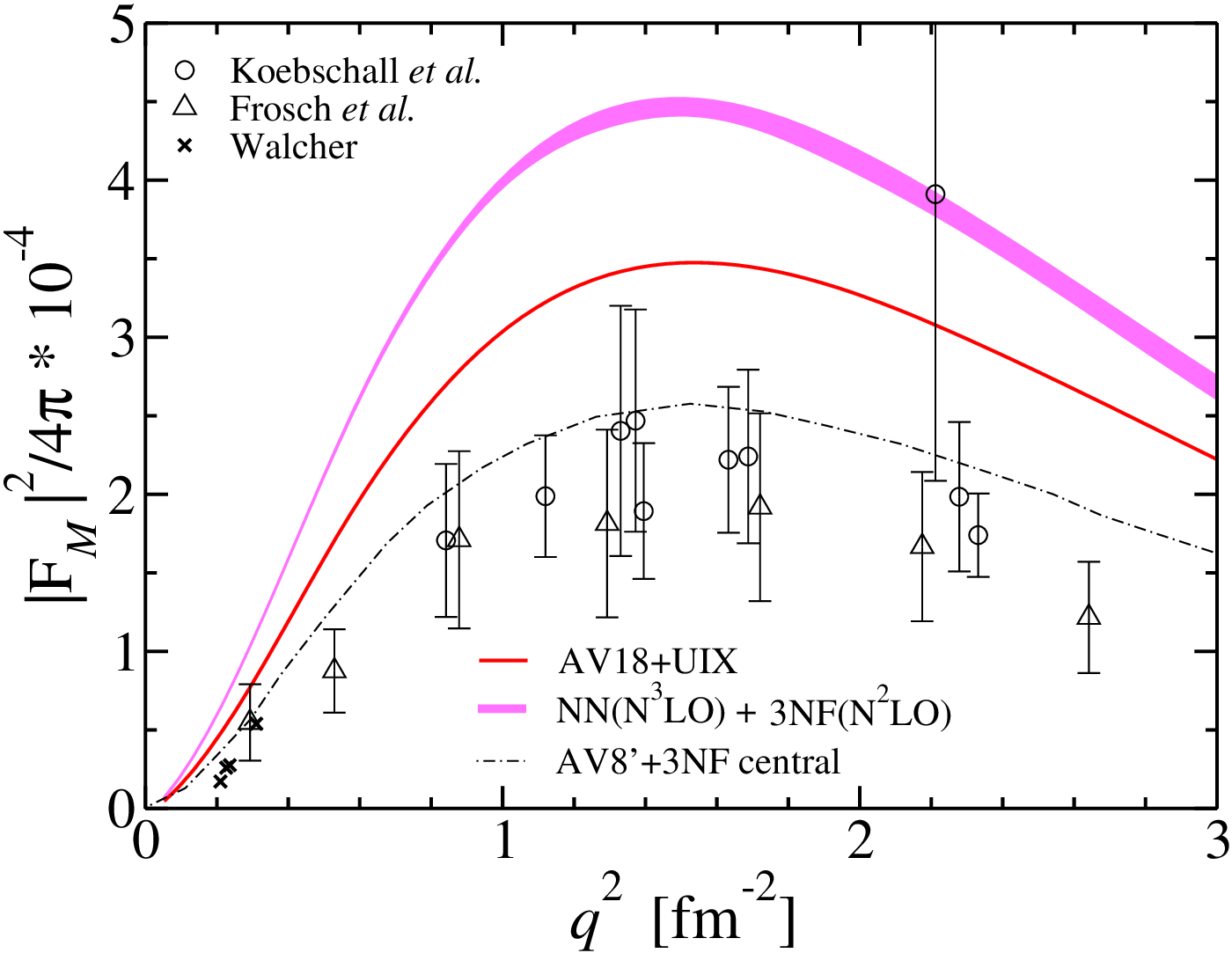}
\caption{Elastic form factor (left) and transition form factor (right) to the first $0^{+}$ excited state of $^4$He with
different three-body Hamiltonians in comparison to the available experimental data. References to the experimental data can be found in~\cite{prism}.
\label{fig_finel}}
\end{figure}

Very recently, we have calculated the  $^4$He  transition form factor $F_{M}(q)$ to the resonant state $0^{+}$~\cite{prism}.  $F_M(q)$ can be measured from electron scattering experiments and several data sets are available.   The first calculation was performed by Hiyama {\it et al.} \cite{Hiyama} with a simple purely central $3NF$ and using bound-state techniques. A good description of data was achieved. We performed a calculation of $F_{M}(q)$ treating the continuum problem with the LIT method and using realistic $3NF$s from traditional nuclear physics~\cite{UIX} and from chiral effective field theory~\cite{EHM}.
We observed a dramatic dependence of the results on the starting three-body Hamiltonian for the transition form factor as shown in Fig.~\ref{fig_finel} (right panel). Three different Hamiltonians (AV18+UIX, $NN$(N$^3$LO)+ $3N$(N$^2$LO) and AV8'+$3NF$ central), which describe the $^4$He ground-state energy to $1\%$ within experiment, show large differences in predicting $F_{M}(q)$.   This is as surprising as interesting and highlights the richness of inelastic observables with
respect to elastic case. In fact, the elastic form factor of the $\alpha$-particle, shown in  the left panel of Fig.~\ref{fig_finel}, is in agreement with data for both realistic forces (AV18+UIX and $NN$(N$^3$LO)+ $3N$(N$^2$LO)). 
The failure of the realistic forces to reproduce the available experimental data for $F_{M}(q)$ is motivating new experimental activity to measure the monopole form factor
 via the  $^4$He($^4$He,$^4$He)$^4$He$^*$ reaction at LNS in Catania with the spectrometer
MAGNEX~\cite{catania}.


\section{Towards Heavier Nuclei}
\label{sec:3}

Continuum calculations of electromagnetic reactions where one starts from interactions that reproduce the $NN$ scattering
phase-shifts are only available for very light nuclei ($2\le A\le 7$). Even though the LIT method has been used in conjunction
with both hyper-spherical harmonics and no core shell model, both methods do not lend them-self to a straightforward application
to the nuclei in the medium-mass range. Coupled-cluster theory, on the contrary, is a very powerful many-body method, which has been successfully applied in the medium-mass regime. By merging the LIT approach with coupled-cluster theory, we have been able to cross the boundaries of the very light nuclei and tackle  the photo-disintegration cross section of $^{16}$O~\cite{cclit}. Photo-nuclear reactions in the medium-mass and heavy nuclei have historically lead to the discovery of the giant dipole resonance, which was first explained as a collective motion of all the protons against the neutrons. Self consistent mean field theories have extensively been applied to the description of the giant dipole resonance, see e.g.~\cite{Erler2011}, and lead to a good agreement with the experimental data, see e.g.~\cite{LyT12}. The effective interactions used in these calculations  are typically calibrated on finite nuclei.

\begin{figure}[htb]
\centering
\sidecaption
\includegraphics[width=7cm,clip]{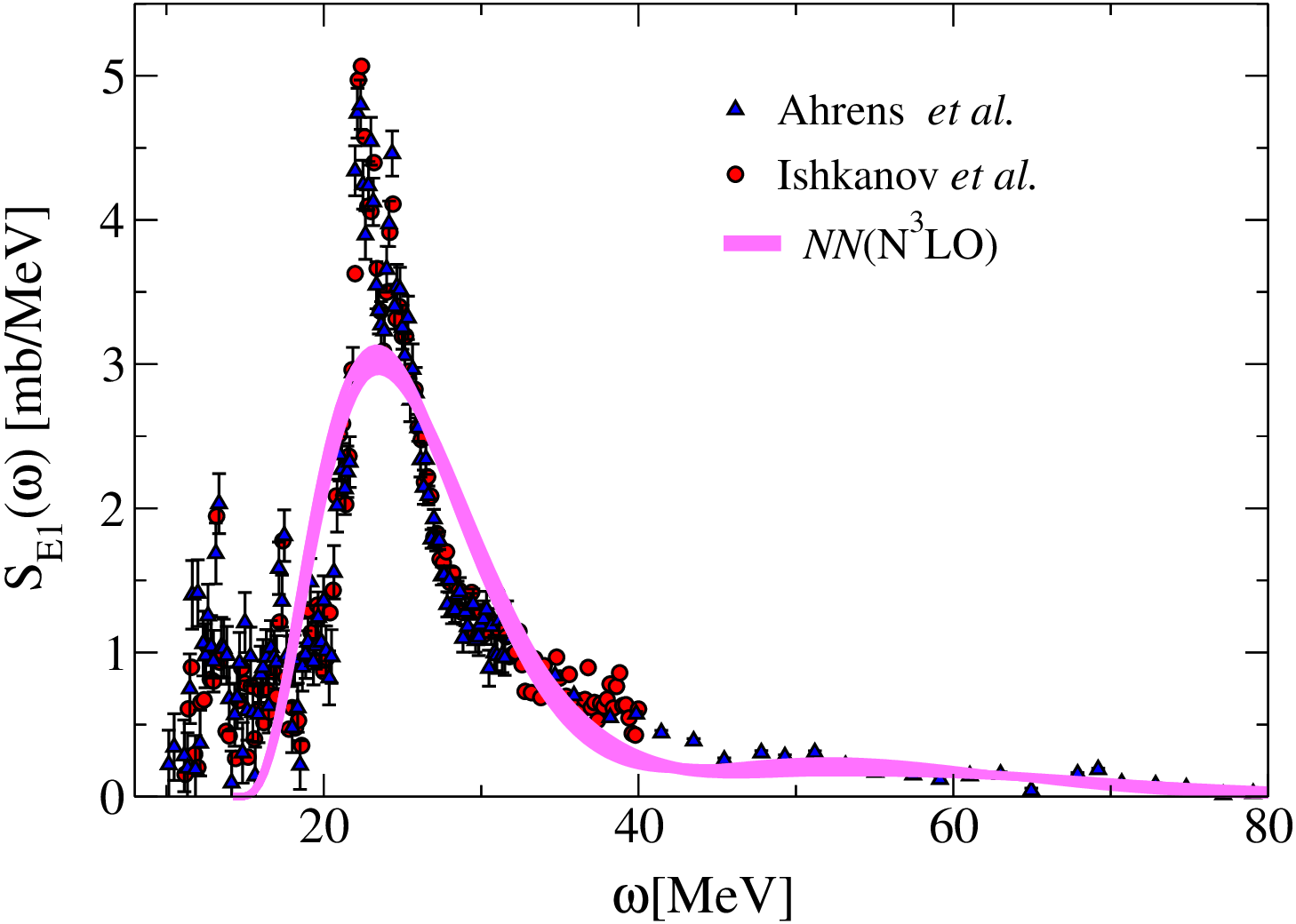}
\caption{Comparison of the calculated $^{16}$O dipole
  response function against experimental data by
  Ahrens {\it et al.}~\cite{ahrens1975} (triangles),
  and Ishkhanov {\it et al.}~\cite{ishkhanov2002} (circles).}
\label{fig-3}       
\end{figure}

We take a different approach and use coupled-cluster theory starting from the $NN$ interaction in chiral
effective field theory at N$^3$LO~\cite{EM}. We  supplement it by a point Coulomb force but omit
the $3NF$s that already appear N$^2$LO.
In Fig.~\ref{fig-3} we present our results for the dipole response function calculated with the single-double approximation scheme
for $^{16}$O. The theoretical prediction is compared to the available experimental data from Ahrens {\it et al.}~\cite{ahrens1975}  and also
to a more recent evaluation by Ishkhanov {\it et al.}~\cite{ishkhanov2002}. 
We observe that the location of the peak and the total strength of the experimental dipole response function are correctly reproduced by our calculation. However, the width of the theoretical resonance is broader than the experimental one.
It is interesting to note that the classical Thomas-Reiche-Kuhn~\cite{TRK} sum
rule  $\left[59.74 \frac{NZ}{A}{\rm MeV~ mb}\right]$ is exhausted by integrating the photo-nuclear cross section up to 40 MeV. When integrating up to 100 MeV we get an enhancement factor of $0.57-0.58$.
The effect of triples, which are neglected in the present calculation, is difficult to estimate. However, we noted that
for the $^4$He case, calculations in the coupled-cluster single-double scheme agree quite nicely with exact hyper-spherical harmonics results from the same potential~\cite{cclit}, indicating that the effect of triples is small.   
Regarding the neglected $3NF$s, it is not clear how large their effect is for $^{16}$O.
The investigation of their impact on the photo-nuclear cross section of medium-mass nuclei is underway.

\section{Conclusions}
\label{sec:5}

Electromagnetic reactions on nuclei are key observables that allow to test our understanding of
the nuclear dynamics via a comparison of accurate ab-initio calculations with experimental data. They also allow to
create strong ties between nuclear physics and other field of physics, like atomic physics or astrophysics.
The prospect of extending ab-initio calculations towards the medium-mass regime is very exciting. The new method
obtained by merging the Lorenz integral transform approach with coupled-cluster theory paves the way for many
future calculations of continuum response functions from first principles. 

\subsection*{Acknowledgments}
I would like to thank all my collaborators for their help in obtaining the results discussed here and
for sharing their insights on these topics.
This work was
supported in part by the Natural Sciences and Engineering Research
Council (NSERC) and the National Research Council (NRC) of Canada.

\end{document}